\documentclass[12pt]{article}
\newcommand{\CR}{\nonumber \\}
\newcommand{\pa}{\partial}
\newcommand{\A}{\alpha}

\newcommand{\G}{\gamma}

\newcommand{\cR}{{\cal R}}

\newcommand{\ep}{\epsilon}

\renewcommand{\thefootnote}{\fnsymbol{footnote}}

\begin{document}
\begin{titlepage}
\begin{flushright}
hep-th/9906023 \\
YITP-99-31 \\
May, 1999
\end{flushright}
\vspace{0.5cm}
\begin{center}
{\Large \bf 
$A$-$D$-$E$ Singularity and the Seiberg-Witten Theory\footnote{
Talk given at the workshop \lq\lq 
Gauge Theory and Integrable Models'', YITP, Kyoto, January 26--29, 1999}
}
\lineskip .75em
\vskip2.5cm
{\large Katsushi Ito}
\vskip 1.5em
{\large\it Yukawa Institute
for Theoretical Physics \\  Kyoto University, Kyoto 606-8502, Japan}  
\vskip 3.5em
\end{center}
\vskip3cm
\begin{abstract}
We study the low-energy effective theory of $N=2$ supersymmetric
Yang-Mills theory with $ADE$ gauge groups in view of the spectral
curves of the periodic Toda lattice and the $A$-$D$-$E$ singularity theory. 
We examine the exact solutions by using the Picard-Fuchs equations for the 
period integrals of the Seiberg-Witten differential.
In particular, we find that  the superconformal fixed point in the strong 
coupling region of the Coulomb branch is characterized by the $ADE$ 
superpotential. 
We compute the scaling exponents, which agree with the previous results. 
\end{abstract}
\end{titlepage}
\baselineskip=0.7cm
\newpage
\renewcommand{\thefootnote}{\arabic{footnote}}
\section{Introduction}
Seiberg and Witten showed that the low-energy effective theory of
$N=2$ supersymmetric gauge theory 
in four dimensions is determined  by the prepotential, a
holomorhic function of the period integrals  of the meromorphic one-form (the 
Seiberg-Witten differential) on a Riemann surface\cite{SeWi}.
For a simple Lie group $G$, it has been proposed in \cite{Go,MaWa} that the 
spectral curve of the periodic Toda lattice associated with the dual
affine Lie algebra $(\widehat{G})^{\vee}$ provides the Riemann surface 
which describes the Coulomb branch of $N=2$ supersymmetric Yang-Mills
theory with the gauge group $G$.
In the case of gauge theories with some matter hypermultiplets, the
spectral curves and related integrable systems are
discussed in \cite{DHPh}.
Other systematic approaches based on the heterotic/type II duality\cite{Le} or
the M5 branes \cite{Wi} are also  studied extensively.

In the preset work we will study the exact solution of the low-energy
effective theory from the viewpoint of the spectral curve of the
periodic Toda lattice.
For $ADE$ type gauge groups, the spectral curves are shown to be 
the sum of the superpotential of two-dimensional topological
Landau-Ginzburg models of $ADE$ type and that of the topological 
 $CP^{1}$ model. 
In a series of papers\cite{ItYa1,ItYa2,ItYa3,ItXiYa}, we have studied 
various aspects of the exact solution of the Seiberg-Witten theories 
with $ADE$ gauge groups by using two-dimensional topological field
theories.

This paper is organized as follows:
In sect. 2, we introduce the spectral curve of the periodic
Toda lattice associated with the dual of the affine Lie algebra
$(\widehat{G})^{\vee}$ for the gauge theory with gauge group $G$.
In sect. 3, we consider the $ADE$ gauge groups and express  
the spectral curve as the sum of the
superpotential of the topological $CP^{1}$ model and the $ADE$
minimal model.
Using the flat coordinates in the $A$-$D$-$E$ singularity theory, 
we derive the Picard-Fuchs differential equations obeyed by  the period
integral of the Seiberg-Witten differential. 
We then show that these equations are equivalent to the Gauss-Manin
system for the $ADE$ minimal model and the $CP^{1}$ model and the
scaling  relation for the Seiberg-Witten differential.
In sect, 4, we study an exact solution in the strong coupling 
region. 
Argyres and Douglas\cite{ArDo} showed that there exists a non-trivial
RG fixed point in the Coulomb branch such that the massless solitons with
mutually non-local charges coexist and  the theory
corresponds to $N=2$ superconformal field theory, where the gauge
invariant order parameters have fractional dimensions with respect to
the BPS mass. 
We investigate this Argyres-Douglas point in the 
Coulomb branch of the $N=2$ supersymmetric Yang-Mills theory for $ADE$ 
gauge groups.

\section{Spectral Curves and $A$-$D$-$E$ Singularity}
The low-energy properties of the Coulomb branch of $N=2$ supersymmetric
gauge theories with gauge group $G$ are exactly described by holomorphic data
associated with certain algebraic curves. 
In particular, the BPS mass formula
is expressed in terms of the period integrals of the so-called
Seiberg-Witten (SW) differential $\lambda_{SW}$:
\begin{equation}
m_{BPS}=| n^{I} a_{I}+m^{I} {a_{D}}_{I}| 
\label{eq:bps}
\end{equation}
where $n^{I}$ and $m^{I}$ are integers and 
\begin{equation}
a_I=\oint_{A_I}\lambda_{SW}, \hskip10mm {a_D}_I=\oint_{B_I}\lambda_{SW},
\quad I=1,\cdots ,r
\label{period}
\end{equation}
along one-cycles $A_I$, $B_I$ with appropriate intersections on the curve.
Here $r$ is the rank of the gauge group $G$.
The low-energy effective theory effective action is described by the 
prepotential ${\cal F}(a)$. 
The dual period ${a_{D}}_{I}$ is then given by $\partial {\cal
F}(a)/\partial a_{I}$.

We define the spectral curve for the periodic
Toda lattice for the (twisted) affine Lie algebra $(\widehat{G})^{\vee}$.
Let $G$ be a simple Lie algebra with rank $r$. 
Let $\A_{1}, \cdots, \A_{r}$ be simple roots of the Lie algebra 
and $\A_{0}=-\theta$, where $\theta$ denotes the highest root. 
We consider a representation ${\cal R}$ with $d$ dimensions. 
Let $E_{\A}$  be generators
associated with the roots $\A$ and $H^{i}$  
those of the Cartan subalgebra, which are realized by $d\times d$
matrices in the representation ${\cal R}$.
Introduce matrices $A$ and $B$ by
\begin{eqnarray}
A(z)&=& \sum_{i=1}^{r} b_{i}H_{i}+a_{i}(E_{\A_{i}}+E_{-\A_{i}})
+a_{0} (z  E_{\A_{0}}+z^{-1}E_{-\A_{0}}) \CR
B(z)&=&  \sum_{i=1}^{r} b_{i}H_{i}+a_{i}(E_{\A_{i}}-E_{-\A_{i}})
+a_{0} (z  E_{\A_{0}}-z^{-1}E_{-\A_{0}}),
\end{eqnarray}
where $z$ is called as the spectral parameter.
The equation of motion of the periodic Toda lattice is defined in
the Lax form
\begin{equation}
{d A\over d t}=\mbox{[} A, B\mbox{]}.
\end{equation}
The spectral curve is defined by the characteristic polynomial of the
matrix $A(z)$:
\begin{equation}
{\cal P}^{\cR}_{G}(x; u_{1},\cdots, u_{r},z)\equiv
{\rm det}_{{\cal R}}(x 1_{d}-A(z))=0.
\label{curve}
\end{equation}
Here $u_{i}$ ($i=1, \cdots, r$)
are the $i$-th order Casimirs of $G$ with degree $q_{i}$ where $q_{i}$ 
is order of the Casimirs. 
The exponents $e_{i}$ 
of $G$ is related to $q_{i}$ by $q_{i}=e_{i}+1$.
The second order Casimir $u_{2}$ has degree 2 and 
the top Casimir $u_{r}$ has degree $q_{r}=h$, where $h$ is the 
Coxeter number. 
The curve depends on the scale parameter
$\mu^{2}=\prod_{i=0}^{r}a_{i}^{n_{i}}$, where non-negative integers
$n_{i}$'s are the Dynkin labels of the affine roots. 
The spectral parameter $z$ and $\mu$ have degree $h^{\vee}$, the dual Coxeter
number of $G$. 
The exponents and the (dual) Coxeter number are given in the table 1.
\begin{table}[h]
\label{tab1}
\begin{center}
\begin{tabular}{lllll}
\hline 
group $G$ & $(\widehat{G})^{\vee}$ & $h$ & $h^{\vee}$ & exponents \\ \hline
$A_{r}$ & $A^{(1)}$  & $r+1$ & $r+1$ & $1,2,\cdots, r$ \\
$B_{r}$ & $A_{2r-1}^{(2)}$ & $2r$ & $2r-1$ & $1,3,\cdots, 2r-1$ \\
$C_{r}$ & $D_{r+1}^{(2)}$ &  $2r$ & $r+1$  & $1,3,\cdots, 2r-1$ \\
$D_{r}$ & $D_{r}^{(1)}$ & $2r-2$ & $2r-2$ & $1,3,\cdots, 2r-3,r-1$ \\
$E_{6}$ & $E_{6}^{(1)}$ & 12 & 12 & 1,4,5,7,8,11 \\
$E_{7}$ & $E_{7}^{(1)}$ & 18 & 18 & 1,5,7,9,11,13,17 \\
$E_{8}$ & $E_{8}^{(1)}$ & 30 & 30 & 1,7,11,13,17,19,23,29 \\
$F_{4}$ & $E_{6}^{(2)}$ & 12 & 9 & 1,5,7,11 \\
$G_{2}$ & $D_{4}^{(3)}$ & 6 & 4 & 1,5 \\
\hline
\end{tabular}
\end{center}
\caption{List of the (dual) Coxeter numbers and exponents}
\end{table}
The spectral curve is invariant under the transformation
$z\rightarrow \mu^{2}/z$. 
For simply laced Lie algebra $G$, we have $h=h^{\vee}$. Therefore the top
Casimir $u_{r}$ and the spectral parameter $z$ or its dual $\mu^{2}/z$ 
have the same degree. 
It is found that for the representation $\cR$ of $G=ADE$, the
spectral curve is given by 
\begin{equation}
P_{G}^{\cR}(x; u_{1}, \cdots, u_{r}+z+{\mu^{2}\over z})=0, 
\end{equation}
where $P^{\cR}_{G}(x;u_{1}, \cdots, u_{r})$  is the characteristic 
polynomial of the representation $\cR$ of $G$ and expressed in the 
form of 
\begin{equation}
P_{G}^{\cR}(x; u_{1}, \cdots, u_{r})=\prod_{i=}^{d}(x-\lambda_{i}\cdot 
a).
\end{equation}
Here $\lambda_{i}$ denote the weight vector of the representation $\cR$.
We list the explicit form of the spectral curve for $A_{r}$,
$D_{r}$ and $E_{6}$ Lie algebras for the $d$-dimensional
representation $\underline{d}$:
\begin{itemize}
 \item $A_{r}$ ($A_{r}^{(1)}, \underline{r+1}$)
\begin{equation}
x^{r+1}-u_{1}x^{r-1}-\cdots -u_{r}-\left(z+{\mu^2\over z}\right)=0
\end{equation}
 \item $D_{r}$ ($D_{r}^{(1)},\underline{2r}$)
\begin{equation}
x^{2r}-u_{1}x^{2r-2}-\cdots-u_{r-2}x^{4}-
u_{r}x^2-u_{r-1}^2-x^{2}\left(z+{\mu^2\over z}\right)=0
\end{equation}
\item
$E_{6}$ ($E_{6}^{(1)}, \underline{27}$) \cite{LeWa}
\begin{equation}
{1\over2} x^{3} \left(z+{\mu^2\over z}+u_{6}\right)^2
-q_{1}(x)  \left(z+{\mu^2\over z}+u_{6}\right)
+q_{2}(x)=0,
\end{equation}
where
\begin{eqnarray}
q_1=&&270 x^{15}+342 u_{1} x^{13}+162 u_{1}^2 x^{11}-252 u_{2} x^{10}
+(26 u_{1}^3+18 u_{3}) x^{9}  \CR
&&  -162 u_{1} u_{2} x^{8}+(6 u_{1} u_{3} -27 u_{4}) x^{7}
-(30 u_{1}^2 u_{2}-36 u_{5}) x^{6}
+(27 u_{2}^2 -9 u_{1} u_{4}) x^{5} \CR
&&  -(3 u_{2} u_{3}-6 u_{1} u_{5}) x^{4}
-3 u_{1} u_{2}^2 x^3-3 u_{2} u_{5} x-u_{2}^3, \CR
q_{2}=&& {1\over 2x^{3}} (q_1^2-p_1^2 p_2), \CR
p_1=&& 78 x^{10}+60 u_{1} x^{8} +14 u_{1}^2 x^{6}-33 u_{2} x^{5}
+2 u_{3} x^{4}-5 u_{1} u_{2} x^{3}-u_{4} x^{2}-u_{5} x-u_{2}^2,  \CR
p_2=&&12 x^{10}+12 u_{1} x^{8}+4 u_{1}^2 x^{6}-12 u_{2} x^{5}+u_{3}x^{4}
-4 u_{1} u_{2} x^{3}-2 u_{4} x^{2}+4 u_{5} x+u_{2}^2. \CR
&& \label{eq:e6}
\end{eqnarray}
\end{itemize}
For simply-laced Lie algebra $G$, $\widehat{G}$ is self-dual, i.e. 
$(G^{(1)})^{\vee}=G^{(1)}$. 
For non-simply laced Lie algebras, 
we have $\widehat{B_{r}}^{\vee}=A_{2r-1}^{(2)}$, 
$\widehat{C_{r}}^{\vee}=D_{r+1}^{(2)}$, 
$\widehat{F_{4}}^{\vee}=E_{6}^{(2)}$ and
$\widehat{G_{2}}^{\vee}=D_{4}^{(3)}$.
Thus we need the twisted affine Lie algebra
to construct the spectral curve. 
The characteristic polynomial can be obtained by folding procedure 
of the corresponding Dynkin diagram. 
Due to $h\neq h^{\vee}$, the spectral parameter $z$ or its
dual $\mu^{2}/z$ appears in the spectral curve in a nontrivial way.
Now we will write the explicit form of the spectral curves for
$d$-dimensional representation $\underline{d}$ of non-simply laced Lie 
algebra $G$.
\begin{itemize}
\item For $B_{r}$ ($A_{2r-1}^{(2)}$, $\underline{2r}$) case, 
the spectral curve of the representation $\underline{2r}$ is obtained
from  the characteristic polynomial 
$P_{A_{2r-1}}^{\underline{2r}}(x; u_{1},\cdots, u_{2r-1})$ by the restriction 
$u_{2}=\cdots=u_{2r-4}=0$ and $u_{2r-2}=z+{\mu^{2}/z}$:
\begin{equation}
x^{2r}-u_{1}x^{2r-2}-\cdots-u_{2r-1}-x\left(z+{\mu^2\over z}\right)=0.
\end{equation}
 \item For $C_{r}$ ($D_{r+1}^{(2)}$:$\underline{2r+2}$)
case, the curve is obtained from the characteristic polynomial 
$
P_{D_{r+1}}^{\underline{2r+2}}(x; u_{1}, \cdots, u_{r}, u_{r+1})
$
by restricting $u_{r}=x-\mu^{2}/z$;
\begin{equation}
x^{2r+2}-u_{1}x^{2r}-\cdots-u_{r-1} x^{4}
-u_{r+1}x^{2}-\left(z-{\mu^2\over z}\right)^2=0.
\end{equation}
 \item For $F_{4}$ ($E_{6}^{(2)}$:$\underline{27}$) case, the curve is 
obtained from the characteristic polynomial 
$
P_{E_{6}}^{\underline{27}}(x; u_{1},u_{2}, u_{3}, u_{4}, u_{5}, u_{6})
$
by the restriction
$u_{2}=0$ and $u_{5}=-6 (z+\mu^{2}/z)$;
\begin{eqnarray}
&& -8 \left( z+{\mu^{2}\over z}\right)^{3} 
+a_{1}(x) \left( z+{\mu^{2}\over z}\right)^{2}
+a_{2}(x) \left( z+{\mu^{2}\over z}\right)
+a_{3}(x)=0, \CR
a_{1}(x) = && \!\!\!\!\!\!\!\!
- 636x^{9} - 300{u_{1}}x^{7} - 48{u_{1}}^{2}x^{5}
 - 5{u_{3}}x^{3} + 2{u_{4}}x,  \CR
a_{2}(x)= && \!\!\!\!\!\!\!\!
- 168 x^{18} - 348 {u_{1}} x^{16} - 276 {u_{1}}^{2} x^{14} 
+ ( - 116 {u_{1}}^{3} + 14 {u_{3}}) x^{12} \CR
 & & \!\!\!\!\!\!\!\!
  + ( - 92 {u_{4}} - 20 {u_{1}}^{4} - 8 {u_{1}} {u
_{3}}) x^{10} + ( - 42 {u_{1}} {u_{4}} - 6 {u_{1}}^{2} {u_{3
}}) x^{8}\CR
&& \!\!\!\!\!\!\!\!
 + ( - 4 {u_{6}} - {\displaystyle \frac {10}{3}}  {u
_{1}}^{2} {u_{4}} - {\displaystyle \frac {2}{3}}  {u_{3}}^{2})
 x^{6} 
  + ({\displaystyle \frac {1}{3}}  {u_{3}} {u_{4}}
 - {\displaystyle \frac {2}{3}}  {u_{6}} {u_{1}}) x^{4}, \CR
a_{3}(x)=&& \!\!\!\!\!\!\!\!
x^{27} + 6 {u_{1}} x^{25} + 15 {u_{1}}^{2} x^{23} + 
(20 {u_{1}}^{3} + {u_{3}}) x^{21} + (5 {u_{4}} + 4 {u_{1}} {u_{3}} + 
15 {u_{1}}^{4}) x^{19}  \CR
 & & \!\!\!\!\!\!\!\!
  + (6 {u_{1}}^{2} {u_{3}} + 12 {u_{1}} {u_{4}} + 
6 {u_{1}}^{5}) x^{17} + ({\displaystyle \frac {1}{3}}  {u_{3}}
^{2} + 5 {u_{6}} + 4 {u_{1}}^{3} {u_{3}} + {\displaystyle 
\frac {26}{3}}  {u_{1}}^{2} {u_{4}} + {u_{1}}^{6}) x^{15} \CR
 & & \!\!\!\!\!\!\!\!
  + ({\displaystyle \frac {4}{3}}  {u_{1}}^{3} {u_{4
}} + {\displaystyle \frac {19}{3}}  {u_{6}} {u_{1}} + {u_{1}}^{
4} {u_{3}} + {\displaystyle \frac {4}{3}}  {u_{3}} {u_{4}} + 
{\displaystyle \frac {2}{3}}  {u_{3}}^{2} {u_{1}}) x^{13} \CR
 & & \!\!\!\!\!\!\!\!
  + ({\displaystyle \frac {1}{3}}  {u_{1}}^{2} {u_{3
}}^{2} - {\displaystyle \frac {1}{3}}  {u_{1}}^{4} {u_{4}} - 
{\displaystyle \frac {15}{4}}  {u_{4}}^{2} + 3 {u_{6}} {u_{1}}
^{2}) x^{11} \CR
 & & \!\!\!\!\!\!\!\!
  + ({\displaystyle \frac {1}{3}}  {u_{6}} {u_{3}}
 - {\displaystyle \frac {4}{9}}  {u_{1}}^{2} {u_{3}} {u_{4}}
 + {\displaystyle \frac {1}{27}}  {u_{3}}^{3} - {\displaystyle 
\frac {13}{6}}  {u_{4}}^{2} {u_{1}} + {\displaystyle \frac {13
}{27}}  {u_{6}} {u_{1}}^{3}) x^{9} \CR
 & & \!\!\!\!\!\!\!\!
  + ( - {\displaystyle \frac {1}{9}}  {u_{3}}^{2} {u
_{4}} - {\displaystyle \frac {1}{2}}  {u_{6}} {u_{4}} + 
{\displaystyle \frac {1}{9}}  {u_{6}} {u_{1}} {u_{3}} - 
{\displaystyle \frac {7}{36}}  {u_{1}}^{2} {u_{4}}^{2}) x^{7}
 + ({\displaystyle \frac {1}{12}}  {u_{4}}^{2} {u_{3}} - 
{\displaystyle \frac {1}{6}}  {u_{6}} {u_{1}} {u_{4}}) x^{5}
 \CR
 & &   + ( - {\displaystyle \frac {1}{54}}  {u_{4}}^{3} - 
{\displaystyle \frac {1}{108}}  {u_{6}}^{2}) x^{3}.
\label{eq:f4}
\end{eqnarray}
\item For $G_{2}$ ($D_{4}^{(3)}$:$\underline{8}$)
case\cite{MaWa}, the spectral curve is obtained from 
$P_{D_{4}}^{\underline{8}}(x; u_{1}, u_{2}, u_{3},
u_{4})
$ by the restriction $u_{1}=2u$,
$u_{2}=-u^{2}-z+\mu^{2}/z$, $u_{3}=\sqrt{3} 
(z-{\mu^{2}/z})$ and $u_{4}=v+2u (z+{\mu^{2}/ z})$;
\begin{equation}
3\left(z-{\mu^{2}\over z}\right)^2
-x^8+2  u x^6-
\left[ u^2 +\left(z+{\mu^{2}\over z}\right)\right] x^4
+\left[ v+ 2 u \left(z+{\mu^{2}\over z}\right) \right] x^2=0.
\end{equation}
\end{itemize}
We may write the spectral curve in the form of 
\begin{equation}
z+{\mu^2 \over z}=W_G^\cR (x,u_1,\cdots ,u_{r}),
\label{eq:solve}
\end{equation}
namely we regard the curve as an fibration over $CP^{1}$ with the
fiber characterized by the function $W^{\cR}_{G}(x)$.
For example, in the case of $A_{r}$, $D_{r}$ and $E_{6}$ gauge groups, 
we obtain
\begin{itemize}
 \item $A_{r}$
\begin{equation}
W_{A_r}^{\underline{r+1}}=x^{r+1}-u_1x^r- \cdots -u_{r-1}x-u_{r},
\end{equation}
 \item $D_{r}$
\begin{equation}
W_{D_r}^{\underline{2r}}=x^{2r-2}-u_1x^{2r-4}- \cdots -u_{r-2}x^2
-{u_{r-1}^2 \over x^2}-u_{r},
\end{equation}
 \item $E_{6}$
\begin{equation}
W_{E_6}^{\underline{27}}={1\over x^3}
 \left( q_1 \pm p_1\sqrt{p_2}\right)-u_6,
\end{equation}
where $p_{1}$, $p_{2}$ and $q_{1}$ are given in (\ref{eq:e6}).
\end{itemize}
It is important to to notice that $W_{ADE}(x)$ is nothing but the
superpotential of the two-dimensional topological Landau-Ginzburg (LG)
model of type $ADE$.
$W_{A_{r}}^{\underline{r+1}}(x)$ and $W_{D_{r}}^{\underline{2r}}(x)$  are
familiar superpotentials for the $A_{r}$ and $D_{r}$ type minimal
models. 
For $E_{6}$ case, the function $W_{E_{6}}^{\underline{27}}$ looks like 
very different from the usual deformation of the $E_{6}$ singularity
written in terms of three variables
\begin{equation}
W_{E_{6}}(x_{1},x_{2},x_{3})=x_{1}^{4}+x_{2}^{3}+x_{3}^{2}.
\label{eq:e6sing}
\end{equation}
It is, however, found in \cite{EY} that $W_{E_{6}}^{\underline{27}}$ 
is a single-variable version of the LG superpotential for the $E_{6}$
minimal model. 
On the other hand, the singularity of the form of (\ref{eq:e6sing})
is obtained by considering the fibration of ALE spaces\cite{Le}. 
The relation of these two description of the Seiberg-Witten curves are
discussed in \cite{LeWa}.

For non-simply laced cases, we have rather nontrivial $W_{G}(x)$
from the spectral curves although in the two dimensional case, the LG
superpotentials are obtained from the simply laced one by the folding
procedure\cite{Zu}. 
The functions $W^{\cR}_{G}(x)$ are given as follows:
\begin{itemize}
\item $B_{r}$
\begin{equation}
W_{B_{r}}^{\underline{2r}}(x;u_{1}, \cdots, u_{r}) 
={W_{BC}(x;u_{1}, \cdots, u_{r})\over x},
\end{equation}
where the LG potential of $BC$ type 
\begin{equation}
W_{BC}(x;u_{1}, \cdots, u_{r})=x^{2r}-\sum_{i=1}^{r} u_{i} x^{2r-2i}
\end{equation}
is obtained from the $A_{2r-1}$ superpotential
$W_{A_{2r-1}}(x;\tilde{u}_{1}, \cdots, \tilde{u}_{2r-1})$ by the restriction 
$\tilde{u}_{2k}=0$
($k=1,\cdots, r-1$) and setting $u_{k}=\tilde{u}_{2k-1}$ ($k=1,\cdots, r$).
\item $C_{r}$
\begin{equation}
W_{C_{r}}^{\underline{2r+2}}(x;u_{1}, \cdots, u_{r}) 
=\left(x^{2}W_{BC}(x;u_{1}, \cdots, u_{r})^{2}+4\mu^{2}\right)^{1/2}.
\end{equation}
\item $F_{4}$
\begin{equation}
W_{F_{4}}^{\underline{27}}={a_{1}(x)\over 24}
-{1\over2}
\left\{ \left( -q+\sqrt{q^{2}+4 p^{3}}\right)^{1/3}
+\left(-q-\sqrt{q^{2}+4 p^{3}}\right)^{1/3}\right\},
\end{equation}
where
\begin{eqnarray}
p(x)&&=-{a_{2}\over6}-{a_{1}^{2}\over 144}, \CR
q(x)&&={1\over27} \left( {a_{1}^{3}\over 32} 
+{9\over8} a_{1} a_{2} +27 a_{3} \right), 
\end{eqnarray}
and $a_{1}$, $a_{2}$ and $a_{3}$ are defined in (\ref{eq:f4}).
\item $G_{2}$
\begin{equation}
W_{G_{2}}^{\underline{8}}={1\over 6} (p_{1}+ \sqrt{p_{1}^2+12 p_{2}}),
\end{equation}
where
\begin{equation}
p_{1}= 6 x^{4}-2 u x^{2}, \hskip10mm 
p_{2}= x^{8} -2 u x^{6}+u^{2} x^{4}-v x^{2} + 12 \mu^{4} .
\end{equation}
\end{itemize}
Note that for $C_{r}$ and $G_{2}$ cases, the superpotentials
$W_{G}^{\cR}(x)$ depend on the scale parameter $\mu$ explicitly.
So far we have seen the SW spectral curves for general
gauge groups. 
The SW differential defined on these spectral curves take
the simple form\cite{MaWa}:
\begin{equation}
\lambda_{SW}={1\over 2\pi i} x {d z \over z}.
\label{eq:swd}
\end{equation}
In the next section we will study the period integrals of the
SW differential using the two-dimensional topological LG
theory. 

\section{Picard-Fuchs Equations and 2D Topological Landau-Ginzburg  Models}
In this section we consider the $ADE$ gauge groups.
To describe the moduli space of the Coulomb branch we adopt the
flat coordinate system $(t_1, t_2, \cdots ,t_r)$ developed 
in the $A$-$D$-$E$ singularity theory instead of the conventional Casimir
coordinates $(u_1, u_2, \cdots ,u_r)$. 
The
coordinate transformation is read off from the residue integral
\begin{eqnarray}
t_i=c_i \oint dx W_G^\cR (x,u)^{e_i \over h}, \hskip10mm i=1,\cdots, r
\label{flat}
\end{eqnarray}
with a suitable constant $c_i$ \cite{EY,EYY}. 
Notice that the overall degree of $W_G^\cR$ is equal to $h$.
The flat coordinates $t_i$ are expressed as polynomials in $u_i$.

Firstly we discuss the role of flat coordinates in 
the two-dimensional topological Landau-Ginzburg
models.
We define the primary fields $\phi_{i}(x,t)$ as the derivatives of the 
superpotential $W_{G}(x,t)$:
\begin{equation}
\phi_{i}(x,t)={\partial W_{G}(x,t)\over \partial t_{i}}.
\end{equation}
We choose the normalization factor $c_{r}$ such that 
$\phi_{r}(x,t)=1$.
We now consider two-dimensional topological gravity coupled to
topological LG model\cite{DVV}. 
In this case,  the primary field $\phi_{r}$ is regarded as the puncture
operator $P$. 
In the flat coordinate system, the topological metric 
$\eta_{ij}=\langle \phi_{i}\phi_{j}P\rangle$ is
independent of $t_{i}$ and takes the form:
\begin{equation}
\eta_{ij}=\delta_{e_{i}+e_{j},h}.
\end{equation}
Furthermore, the primary fields obey the operator product expansions
\begin{equation}
\phi_i (x,t)\, \phi_j (x,t)=\sum_{k=1}^r {C_{ij}}^k(t)\, \phi_k (x,t)
+Q_{ij}(x,t) \, \partial_x  W_{G}(x,t).
\label{eq:ope}
\end{equation}
The flatness condition implies that the function $Q_{ij}(x,t)$ in 
(\ref{eq:ope}) satisfies
\begin{equation}
{\partial^2 W_{G}(x,t) \over \partial t_i \partial t_j}
=\partial_x \, Q_{ij}(x,t).
\end{equation}
The structure constants $C_{ij}^{k}(t)$ are related to the three-point 
function $F_{ijk}(t)=\langle \phi_{i}\phi_{j}\phi_{k}\rangle$ by the
relation $F_{ijk}(t)=\eta_{kl}C_{ij}^{l}(t)$. 
In two-dimensional topological theory, all the topological correlation 
functions are determined by the free energy $F(t)$.
The three point function $F_{ijk}(t)$ is given by 
$\pa^{3} F(t)/\pa t_{i}\pa t_{j} \pa t_{k}$.

The associativity of the chiral ring 
$(\phi_{i}\phi_{j})\phi_{k}=\phi_{i}(\phi_{j}\phi_{k})$ implies the
relation
$C_{ij}^{l} C_{l k}^{n}=C_{j k}^{l} C_{i l}^{n}$.
Let us introduce $r\times r$ matrices $C_{i}$, $F_{i}$ and $\eta$ by 
$(C_{i})_{j}^{k}=C_{ij}^{k}$,  $(F_{i})_{jk}=F_{ijk}$ and $\eta=(\eta_{ij})$. 
The associativity condition leads to the commutativity of the matrices
$C_{i}$;
\begin{equation}
C_{i} C_{j}=C_{j} C_{i}.
\end{equation}
Using $F_{i}=\eta C_{i}$, we obtain the 
Witten-Dijkgraaf-Verlinde-Verlinde (WDVV) equation:
\begin{equation}
F_{i}\eta^{-1} F_{j}=F_{j}\eta^{-1} F_{i},
\label{eq:wdvv}
\end{equation}
which is one of the important relations in two-dimensional topological 
theory.

Now we go back to the four-dimensional gauge theory. 
We consider the period integral of the SW differential 
$\lambda_{SW}$ (\ref{eq:swd}):
\begin{equation}
\Pi=\int_{\G}\lambda_{SW}
\end{equation}
along the certain one-cycle $\G$ on the spectral curve (\ref{eq:solve}).
In terms of the superpotential $W_{G}(x,t)$, the SW
differential  takes the form:
\begin{equation}
\lambda_{SW}={1\over2\pi i} {x\pa_{x}W_{G}(x,t)\over
 \sqrt{W_{G}(x,t)^2-4\mu^2}} d x.
\end{equation}
It is shown in \cite{ItYa1} that $\Pi$ obeys the set of 
differential equations:
\begin{equation}
\pa_{t_{i}}\pa_{t_{j}}\Pi=\sum_{k=1}^{r} C_{ij}^{k}(t)
\pa_{t_{k}}\pa_{t_{r}}\Pi,
\label{eq:gm}
\end{equation}
which is called as the Gauss-Manin system in the singularity theory. 

In addition to the Gauss-Manin system, the SW differential satisfies 
another two differential equations. 
{}From the scaling relation to the superpotential
\begin{equation}
\left(\sum_{i=1}^{r}q_{i}t_{i}\pa_{t_{i}}+x\pa_{x}\right) W_{G}(x,t)
=h W_{G}(x,t), 
\end{equation}
we obtain the scaling relation for the period $\Pi$:
\begin{equation}
\left(\sum_{i=1}^{r}q_{i}t_{i}\partial_{t_{i}}+h^{\vee}\mu\partial_{\mu}
-1\right)\Pi=0.
\label{eq:scale}
\end{equation}
The final differential equation is obtained by regarding the l.h.s. of
the spectral curve (\ref{eq:solve}) as the superpotential of the
topological $CP^{1}$ model \cite{cp1}:
\begin{equation}
W_{CP^{1}}(z)=z+{\mu^{2}\over z}-t_{r}.
\end{equation}
Since $\log \mu^{2}$ and $t_{r}$ are flat coordinates of the $CP^{1}$
model, we obtain the $CP^{1}$ relation:
\begin{equation}
\left( (\mu\pa_{\mu})^2 -4 \mu^{2} \pa_{t_{r}}^{2}\right) 
\Pi=0 . 
\label{eq:cpone}
\end{equation}
Combining the scaling relation (\ref{eq:scale}) and the $CP^{1}$
relation (\ref{eq:cpone}), we obtain the differential equation 
\begin{equation}
\left\{\left( \sum_{i=1}^{r}q_{i}t_{i}{\pa\over \pa t_{i}}-1 \right)^2
-4\mu^2 h^{2}{\pa^{2}\over \pa t_{r}^2} \right\}\Pi=0
\label{eq:scaling}
\end{equation} 
By solving the Gauss-Manin system (\ref{eq:gm}) and the scaling
equation (\ref{eq:scaling}), we may analyze the exact solutions in the 
Coulomb branch. 
For classical gauge groups, the present Picard-Fuchs equations are
shown to be the same as those appeared in the previous
works\cite{pf1}, in which various gauge theories with or without 
hypermultiplets are discussed.

In the weak coupling region where the scale parameter
$\Lambda^{2h}=4\mu^{2}$ are small, the solutions of these Picard-Fuchs 
equations are studied extensively using various methods\cite{pf1,weak},
which are shown to agree with the microscopic instanton calculation
\cite{micro}. 
In the next section, we study the solutions in the strong coupling
region. 
But before going to the next section, we discuss an important consequence
of the Gauss-Manin system. 
Since the dual period ${a_{D}}_{I}$ also satisfies the Gauss-Manin
system (\ref{eq:gm}), this Gauss-Manin system 
 provides the third-order differential equation 
for the prepotential ${\cal F}(a)$:
\begin{equation}
\widetilde{\cal F}_{ijk}=\sum_{l=1}^{r} {C_{ij}}^{l}\widetilde{\cal
F}_{lrk},
\label{eq:wdvvsw1}
\end{equation}
where 
\begin{eqnarray}
\widetilde{\cal F}_{ijk}&=&\partial_{t_{i}}a_{I} \partial_{t_{j}}a_{J}
\partial_{t_{k}}a_{K} {\cal F}_{IJK}, \CR
{\cal F}_{IJK}&=&\partial_{a_{I}}\partial_{a_{J}}\partial_{a_{K}}{\cal F}(a)
\end{eqnarray}
The equations (\ref{eq:wdvvsw1}) is very similar to
$F_{ijk}(t)=C_{ij}^{l}\eta_{kl}$ in two-dimensional topological
theory. 
Let us introduce matrices $\widetilde{\cal F}_{i}$, ${\cal G}$ and
${\cal F}_{I}$ 
defined by $(\widetilde{\cal F}_{i})_{jk}=\widetilde{\cal F}_{ijk}$, 
 ${\cal G}=\widetilde{\cal F}_{r}$, and 
$({\cal F}_I)_{JK}={\cal F}_{IJK}$, respectively. 
Then we find the WDVV equations in the Seiberg-Witten
theory\cite{wdvv,ItYa3}:
\begin{equation}
\widetilde{\cal F}_{i} {\cal G}^{-1} \widetilde{\cal F}_{j}
= \widetilde{\cal F}_{j} {\cal G}^{-1} \widetilde{\cal F}_{i}, 
\end{equation}
which may be written in the form 
\begin{equation}
{\cal F}_{I} {\cal F}_{K} ^{-1} {\cal F}_{J}
= {\cal F}_{J} {\cal F}_{K}^{-1} {\cal F}_{I}. 
\end{equation}

In addition to the WDVV equation, the prepotential satisfies the
scaling equation\cite{scale}:
\begin{equation}
\left(\sum_{I=1}^{r}a_{I}\pa_{a_{I}}+h^{\vee}\mu\pa_{\mu}\right)
{\cal F}(a)=2 {\cal F}(a).
\end{equation}
This scaling equation is important to calculate the instanton
correction to the prepotential in the weak coupling region\cite{ItYa2}.
Recently it is shown that the prepotential satisfies further
non-trivial equations \cite{RG} obtained from the Whitham hierarchy.
Instanton correction to the prepotential has been calculated in this
framework\cite{EdGoMa} for some gauge theories. 
It would be interesting to study the relation between the WDVV
equation approach and this formulation.

\section{Superconformal point}
One of interesting phenomena in the strong coupling physics of $N=2$
supersymmetric gauge theory is the existence of non-trivial $N=2$  
superconformal fixed point in the Coulomb branch\cite{ArDo,ArPlSeWi,EgHoItYa}. 
At this point, massless solitons of mutually nonlocal charges coexist.
The superconformal field theory is characterized by the scaling
operators with non-trivial (fractional) scaling dimensions. 
In particular, the calculations of the scaling exponents based on the
exact solution suggest that the superconformal fixed points are
characterized by the $A$-$D$-$E$ classification\cite{EgHoItYa}. 
We will study this RG fixed point in view of the Picard-Fuchs
equations obtained in the previous section.

For $G=ADE$, the superconformal fixed points exist at 
$$
t_{1}=\cdots =t_{r-1}=0, \quad t_{r}=\pm 2\mu.
$$
We take the plus sign without loss of generality. 
Let us introduce new flat coordinates $\tilde{t}_{i}$ by 
shifting $t_{r}$ by $2\mu$:
\begin{equation}
t_{i}=\tilde{t}_{i}, \quad (i=1,\cdots,r-1), \quad 
t_{r}=\tilde{t}_{r}+2\mu 
\end{equation}
Since the OPE coefficients $C_{ij}^{k}(t)$ are independent of
$t_{r}$ \cite{ItYa2},  
the Gauss-Manin system does not change its form under this change of
coordinates: 
\begin{equation}
\left( \partial_{\tilde{t}_{i}}\partial_{\tilde{t}_{j}}
-\sum_{k=1}^{r} {C_{ij}}^{k}(\tilde{t})
\partial_{\tilde{t}_{k}}\partial_{\tilde{t}_{r}}\right)\Pi=0,
\label{eq:gmscf}
\end{equation}
The scaling equation (\ref{eq:scaling}), on the other hand, becomes
\begin{equation}
\left\{\left( \sum_{i=1}^{r}q_{i}\tilde{t}_{i}\pa_{\tilde{t}_{i}}-1 \right)^2
+4\mu h 
\left[
\left( \sum_{i=1}^{r}q_{i}\tilde{t}_{i}\pa_{\tilde{t}_{i}}-1 \right)
\pa_{\tilde{t}_{r}}
+{h\over2} \pa_{\tilde{t}_{r}}
\right] \right\}\Pi=0.
\label{eq:scascf}
\end{equation}
Introduce the scaling parameter $\ep$ by 
\begin{equation}
\tilde{t}_{i}=\ep^{q_{i}}\rho_{i}, \quad (i=1,\cdots,r-1), \quad
\tilde{t}_{r}=\ep^{h}
\end{equation}
and consider the limit $\ep\rightarrow 0$. 
We are interested in the solution of  the SW periods which  behave like
\begin{equation}
\Pi=\ep^{\A} f(\rho) +\cdots,
\label{eq:scfsol}
\end{equation}
as $\ep\rightarrow 0$.
Since 
\begin{equation}
\pa_{\tilde{t}_{i}}=\ep^{-q_{i}}\pa_{\rho_{i}}, \quad 
\pa_{\tilde{t}_{r}}={1\over h} \ep^{-h} \left(
\ep\pa_{\ep}-\sum_{i=1}^{r-1}q_{i}\rho_{i}\pa_{\rho_{i}}
\right), 
\end{equation}
the Gauss-Manin system (\ref{eq:gmscf}) for $i,j<r$ becomes
the differential equations for $f(\rho)$ with respect to $\rho$:
\begin{eqnarray}
&& \left( \partial_{\rho_{i}}\partial_{\rho_{j}}
-{1\over h} \sum_{k=1}^{r-1} {\tilde{C}_{ij}}^{k}(\rho)
\partial_{\rho_{k}}
\left[
\A-\sum_{l=1}^{r-1}q_{l}\rho_{l}\pa_{\rho_{l}}
\right] \right. \CR
&& \left. -{1\over h^{2}}  {\tilde{C}_{ij}}^{r}(\rho)
\left[
\A-h-\sum_{l=1}^{r-1}q_{l}\rho_{l}\pa_{\rho_{l}}
\right]
\left[
\A-\sum_{l=1}^{r-1}q_{l}\rho_{l}\pa_{\rho_{l}}
\right]
\right)f(\rho)=0
\end{eqnarray}
where
$
\tilde{C}_{ij}^{k}(\rho)=C_{ij}^{k}(\tilde{t})\ep^{q_{i}+q_{j}-q_{k}-h}.
$
As for the scaling equation (\ref{eq:scascf}), the second term is 
dominant in the superconformal limit. 
We thus find that $f(\rho)$ should satisfy the equation
\begin{eqnarray}
\left(\A-{h+2\over2}\right)
\left[
\A-\sum_{l=1}^{r-1}q_{l}\rho_{l}\pa_{\rho_{l}}
\right]
f(\rho)=0
\end{eqnarray}
This equation determines the exponent $\A$ in (\ref{eq:scfsol}) such as 
\begin{equation}
\A={h+2\over2}.
\label{eq:exp}
\end{equation}
The superconformal field theory is characterized by the scaling
operator ${\rm tr}\phi^{q_{i}}$, whose conformal dimension is 
measured with respect to the BPS mass (\ref{eq:bps}). 
{}From (\ref{eq:scfsol}) and (\ref{eq:exp}), we have 
\begin{equation}
\langle {\rm tr}\phi^{q_{i}}\rangle
\sim (m_{BPS})^{{2q_{i}\over h+2}} 
\end{equation}
Thus the scaling  dimension of $\langle {\rm tr} \phi^{q_{i}}\rangle$
is $2q_{i}/(h+2)$, which agrees with the result of \cite{EgHoItYa}.

We could examine the above arguments from the viewpoint of the SW curve.
In terms of the parameters $\tilde{t}$, the SW curve becomes
\begin{equation}
z+{\mu^{2}\over z}=W_{G}(x; t)=W_{G}(x; \tilde{t})-2\mu. 
\end{equation}
Let us introduce $\xi$ by
\begin{equation}
\xi=\sqrt{z}+{\mu\over \sqrt{z}}.
\end{equation}
Then the curve is expressed in the form of the $ADE$ superpotential with 
the Gaussian part $\xi^{2}$:
\begin{equation}
\xi^{2}=W_{G}(x; \tilde{t}) .
\end{equation}
The SW differential then becomes 
\begin{equation}
\lambda_{SW}={1\over2\pi i} x{dz\over z}
={1\over2\pi i} 2 x {d \xi \over \sqrt{\xi^{2}-4 \mu}}.
\end{equation}
In the superconformal limit, expanding  the above formula around $\ep=0$
we obtain 
\begin{equation}
\lambda_{SW}=-{1\over2\pi i} {1 \over \sqrt{-\mu}}\sum_{n=0}^{\infty} 
{(2n)!\over 2^{2n} (n!)^{2}} {W_{G}(x,\tilde{t})^{{2n+1\over2}} d x \over
(2n+1) (4\mu)^{n}},
\label{eq:swexpa}
\end{equation}
up to the total derivative term.
After rescaling $x=\ep\tilde{x}$, the leading term in (\ref{eq:swexpa})
is 
\begin{equation}
\lambda_{SW}=-{1\over2\pi i  \sqrt{-\mu}} \ep^{h+2\over2}
\sqrt{W_{G}(\tilde{x};\rho_{1},\cdots, \rho_{r-1},1)} d\tilde{x}+\cdots  ,
\end{equation}
which also leads to the exponent (\ref{eq:exp}).
Note that the derivative of the SW period integrals:
$$
{\pa \Pi\over \pa \rho_{i}}\sim
\ep^{h+2\over2} \int {\pa_{\rho_{i}}W_{G}(\tilde{x};\rho)\over 
\sqrt{W_{G}(\tilde{x};\rho)}} d\tilde{x}
$$
is the period of the curve 
\begin{equation}
y^{2}=W_{G}(\tilde{x};\rho_{1},\cdots, \rho_{r-1},1).
\label{eq:swhe}
\end{equation}
Thus we find that the superconformal fixed point is simply
characterized by the $ADE$ superpotential $W_{G}(x,\tilde{t})$.
The SW curve reduce to the hyperelliptic type curve (\ref{eq:swhe}). 
For example, let us consider the $A_{2}$ case. 
The SW curve (\ref{eq:swhe}) is given by 
$$
y^{2}=x^{3}-\rho x-1,
$$
which is nothing but the curve of $SU(2)$ gauge theory with $N_{f}=1$
at the superconformal point (the small torus in \cite{ArDo}). 
The Gauss-Manin  system (\ref{eq:gmscf}) becomes
\begin{equation}
\left[ 
(4\rho^{3}-27)\pa_{\rho}^{2}-{5\over4}\rho
\right] f(\rho)=0.
\end{equation}
One may solve this equation around $\rho=0$ found that the results
agree with those obtained in \cite{ArDo}.

\section{Discussion}
In this paper, we have seen the close relationship between the
four-dimensional gauge theory with $ADE$ gauge group and 
two-dimensional topological LG models coupled to topological gravity. 
We have examined the exact solutions around the superconformal points 
using the Picard-Fuchs equations and showed that the superconformal points
are simply characterized by the $A$-$D$-$E$ singularity. 
For other gauge theories with or without matter hypermultiplets, on
the other hand, it is difficult to extend the present results because
of the complexity of the superpotential. 
But at the superconformal point we would expect that the curves become
simple and are classified by the $A$-$D$-$E$ singularity\cite{EgHoItYa}.
The two-dimensional topological field theory would provide a
useful tool for analyzing the physics in both weak and strong coupling 
region. 

It seems interesting to compare the expansion (\ref{eq:swexpa}) of the
SW differential around the superconformal point with that around the
origin ($t_{i}=0$) of the Coulomb branch:
\begin{equation}
\lambda_{SW}=-{1\over2\pi i} {1\over \sqrt{-4\mu^{2}}}
\sum_{n=0}^{\infty} 
{(2n)!\over 2^{2n} (n!)^{2}} {W_{G}(x,t)^{2n+1} d x \over
(2n+1) (4\mu^{2})^{n}}.
\label{eq:oexpa}
\end{equation}
In \cite{ItXiYa}, it has been shown that the period integrals of
(\ref{eq:oexpa}) are expressed directly in terms of 
the one-point function $\langle \sigma_{n}(\phi_{i})\rangle$ 
of the $n$-th gravitational descendant
$\sigma_{n}(\phi_{i})$ of the primary field $\phi_{i}$
in two-dimensional topological LG models coupled to topological
gravity.
The one-point function  satisfies the
same Gauss-Manin system, which is evaluated by the residue 
integrals\cite{EYY}:
\begin{equation}
\langle \sigma_{n}(\phi_{i})\rangle=b_{n,i} \sum_{j=1}^{r}
\eta_{ij}\oint W_{G}(x,t)^{{e_{j}\over h}+n+1},
\end{equation}
where $b_{n,i}$ is certain constant.
In this formulation, the Gauss-Manin system (\ref{eq:gm}) is also 
derived from the topological
recursion relation \cite{WiDi}:
\begin{equation}
\langle \sigma_{n}(\phi_{i}) X Y \rangle = \sum_{j}
\langle \sigma_{n-1}(\phi_{i})\phi_{j}\rangle \langle \phi^{j} X Y \rangle
\end{equation}
and the puncture equation (in the small phase space):
\begin{equation}
\langle P \prod_{i=1}^{s} \sigma_{n_{i}}(\phi_{\A_{i}})\rangle
=\sum_{i=1}^{s} \langle \prod_{j=1}^{s}\sigma_{n_{j}-\delta_{ji}}
 (\phi_{\A_{j}})\rangle.
\end{equation}
One might expect that the similar relation would be hold in the 
superconformal case, which would be important to understand the
relation between the SW theory and  $d<1$ topological strings.

\vskip3mm\noindent
{\bf Acknowledgments} \ 
The author would like to thank C.-S. Xiong and S.-K. Yang for useful 
discussions.
This work is supported in part by 
the Grant-in-Aid from the Ministry of Education, Science and Culture,
Priority Area: \lq\lq Supersymmetry and Unified
Theory of Elementary Particles'' (\#707).

\newpage


\begin{thebibliography}{99}
\bibitem{SeWi} N. Seiberg and E. Witten
Nucl. Phys. {\bf B426} (1994) 19, {\tt hep-th/9407087}; 
 Nucl. Phys. {\bf B431} (1994) 484,
 {\tt hep-th/9408099}

\bibitem{Go} A. Gorskii, I. Krichever, A. Marshakov, A. Mironov and
A. Morozov, Phys. Lett. {\bf B355} (1995) 466, {\tt hep-th/9505035}

\bibitem{MaWa} E.J. Martinec and N.P. Warner, Nucl. Phys. {\bf B459}
 (1996) 97, {\tt hep-th/9509161}

\bibitem{DHPh}
R. Donagi and E. Witten, Nucl. Phys. {\bf B460} (1996) 299,
{\tt hep-th/9510101} \\
H. Itoyama and A. Morozov, Nucl. Phys. {\bf B491} (1997) 529,
{\tt hep-th/9512161} \\
A. Gorsky, A. Marshakov, A. Mironov, A. Morozov, Phys. Lett. {\bf
B380} (1996) 75, {\tt hep-th/9603140} \\
E. D'Hoker and D.H. Phong, Nucl. Phys. {\bf B513} (1998) 405,
{\tt hep-th/9709053};
Nucl. Phys. {\bf B534} (1998) 697, {\tt hep-th/9804126} \\
A.J. Bordner, E. Corrigan and  R. Sasaki, Prog. Theor. Phys. {\bf 100}
(1998) 1107, {\tt hep-th/9805106} \\
A. J. Bordner, R. Sasaki and K. Takasaki, Prog. Theor. Phys. {\bf 101} 
(1999) 487, {\tt hep-th/9809068} \\
A. Gorsky and A. Mironov, {\it Solutions to the reflection equation
and integrable systems for $N=2$ SQCD with classical groups},
{\tt hep-th/9902030}

\bibitem{Le}
A. Klemm, W. Lerche, P. Mayr, C. Vafa and N. Warner,
Nucl. Phys. {\bf B477} (1996) 746, {\tt hep-th/9604034} \\
For reviews, see W. Lerche, {\it Introduction to Seiberg-Witten Theory 
and its Stringy Origin}, {\tt hep-th/9611190} \\
A. Klemm, {\it On the Geometry behind $N=2$ Supersymmetric Effective
Actions in Four Dimensions}, {\tt het-th/9705131}

\bibitem{Wi}
E. Witten, Nucl. Phys. {\bf B500} (1997) 3, {\tt hep-th/9703066}

\bibitem{ItYa1}
K. Ito and S.-K. Yang, Phys. Lett. {\bf B415} (1997) 45,
{\tt hep-th/9708017}

\bibitem{ItYa2}
K. Ito and S.-K. Yang, Int. J. Mod. Phys. {\bf A13} (1998) 5373,
{\tt hep-th/9712018} 

\bibitem{ItYa3}
K. Ito and S.-K. Yang, Phys. Lett. {\bf B433} (1998) 58, 
{\tt hep-th/9803126}

\bibitem{ItXiYa}
K. Ito, C.-S. Xiong and S.-K. Yang, Phys. Lett. {\bf B441} (1998) 155,
{\tt hep-th/9807183}


\bibitem{ArDo}
P. Argyres and M. Douglas, Nucl. Phys. {\bf B448} (1995) 93,
{\tt hep-th/9505062} 


\bibitem{LeWa}
W. Lerche and N.P. Warner, Phys. Lett. {\bf B423} (1998) 79, 
{\tt hep-th/9608183}

\bibitem{EY} T. Eguchi and S.-K. Yang, Phys. Lett. {\bf B394} (1997)
315, 
{\tt hep-th/9612086}


\bibitem{Zu} 
J.B. Zuber, Mod. Phys. Lett. {\bf A9} (1994) 749,
{\tt hep-th/9312209}

\bibitem{EYY}
 T. Eguchi, Y. Yamada and S.-K. Yang, Mod. Phys. Lett. 
{\bf A8} (1993) 1627

\bibitem{DVV}
R. Dijkgraaf, H. Verlinde and E. Verlinde, Nucl. Phys. {\bf B352}
(1991) 59

\bibitem{WiDi}
E. Witten, Nucl. Phys. {\bf B340} (1989); 
Surv. Diff. Geom. {\bf 1} (1991) 243; \\
R. Dijkgraaf and E. Witten, Nucl. Phys. {\bf B342} (1990) 486

\bibitem{cp1}
B. Dubrovin, {\it Geometry of 2D Topological Field Theories},
{\tt hep-th/9407018}; \\
T. Eguchi, K. Hori and C.-S. Xiong,
Int. J. Mod. Phys. {\bf A12} (1997) 1743, 
{\tt hep-th/9605225}; \\
A. Givental, {\it Stationary Phase Integrals, Quantum Toda Lattices,
Flag Manifolds and the Mirror Conjecture}, 
{\tt alg-geom/9612001}; \\
A. Takahashi, {\it Primitive Forms, Topological LG Models Coupled 
to Gravity and Mirror Symmetry}, 
{\tt math.AG/9802059}

\bibitem{pf1}
A. Klemm, W. Lerche and S. Theisen, Int. J. Mod. Phys. {\bf A11} (1996)
1929, 
{\tt hep-th/9505150}\\
K.~Ito and S.-K.~Yang,   Phys. Lett. {\bf B366} (1996) 165, 
{\tt hep-th/9507144};
in \lq\lq Frontiers In Quantum Field Theory'' p. 331
(World Scientific, Singapore,1996), {\tt hep-th/9603073} \\
M.~Matone, Phys. Lett. {\bf B357} (1995) 342, {\tt hep-th/9506102} \\
Y.~Ohta, J. Math. Phys. {\bf 37} (1996) 6074, {\tt hep-th/9604051};
 {\bf 38} (1997) 682, {\tt hep-th/9604059}; 
 {\bf 40} (1999) 1891, {\tt hep-th/9809180};
{\it Picard-Fuchs Ordinary Differential Systems in $N=2$
Supersymmetric Yang-Mills Theories}, {\tt hep-th/9812085}\\
K.~Ito and N.~Sasakura, Nucl. Phys. {\bf B484} (1997) 141,
{\tt hep-th/9608054} \\
H.~Ewen, K.~Foerger and S.~Theisen, Nucl. Phys. {\bf B485} (1997) 63,
{\tt hep-th/9609062}  \\
H.~Ewen and  K.~Foerger, Int. J. Mod. Phys. {\bf A12} (1997) 4725, 
{\tt hep-th/9610049}; \\
J.M. Isidro, A. Mukherjee, J.P. Nunes and  H.J. Schnitzer, 
Nucl .Phys. {\bf B492} (1997) 647, {\tt hep-th/9609116}; 
Int. J. Mod. Phys. {\bf A13} (1998) 233, {\tt hep-th/9703176}; 
Nucl. Phys. {\bf B502} (1997) 363, {\tt hep-th/9704174}; \\
M. Alishahiha, Phys. Lett. {\bf B398} (1997) 100, {\tt hep-th/9609157}; 
Phys. Lett. {\bf B418} (1998) 317, {\tt hep-th/9703186} \\
K. Ito, Phys. Lett. {\bf B406} (1997) 54, {\tt hep-th/9703180} \\
A.M. Ghezelbash and A. Shafiekhani, Mod. Phys. Lett. {\bf A13} (1998)
527, {\tt hep-th/9708073}; \\
A.M. Ghezelbash, Phys. Lett. {\bf B423} (1998) 87, {\tt hep-th/9710068}

\bibitem{weak}
E.~D'~Hoker, I.M.~Krichever and D.H.~Phong, 
Nucl. Phys. {\bf B489} (1997) 179 {\tt hep-th/9609041};
{\bf B489} (1997) 211 {\tt hep-th/9609145} \\
T. Masuda and H. Suzuki, Int. J. Mod. Phys. {\bf A12} (1997) 3413,
{\tt hep-th/9609006}; Int. J. Mod. Phys. {\bf A13} (1998) 1495,
{\tt hep-th/9609065}; Nucl. Phys. {\bf B495} (1997) 149, {\tt
hep-th/9612240}  \\
T. Masuda, T. Sasaki and H. Suzuki, Int. J. Mod. Phys. {\bf A13}
(1998) 3121, {\tt hep-th/9705166}

\bibitem{micro}
D.~Finnell and P.~Pouliot, Nucl. Phys.  {\bf B453} (1995) 225, 
{\tt hep-th/9503115}; \\
H.~Aoyama, T.~Harano, M.~Sato and S.~Wada, Phys. Lett. {\bf B388} (1996)
331, {\tt hep-th/9607076}\\
K.~Ito and N.~Sasakura, Phys. Lett. {\bf B382} (1996) 95, {\tt hep-th/9602073};
Nucl. Phys. {\bf B484} (1997) 141, {\tt hep-th/9608054}; 
Mod. Phys. Lett. {\bf A12} (1997) 205, {\tt hep-th/9609104} \\
T.~Harano and M.~Sato, Nucl .Phys. {\bf B484} (1997) 167, {\tt
hep-th/9608060}  \\
N.~Dorey, V.V.~Khoze and M.P.~Mattis, Phys. Rev. {\bf D54} (1996)
2921, {\tt hep-th/9603136}; 7832, {\tt hep-th/9607202}; 
Phys. Lett. {\bf B388} (1996) 324, {\tt hep-th/9607066}; 
 {\bf B390} (1997) 205, {\tt hep-th/9606199}; 
{\bf B396} (1997) 141, {\tt hep-th/9611016};
Nucl. Phys. {\bf B492} (1997) 607; \\
F.~Fucito and G.~Travaglini, Phys. Rev. {\bf D55} (1997) 1099,
{\tt hep-th/9605215} \\
A.~Yung, Nucl. Phys. {\bf B485} (1997) 38, {\tt hep-th/9605096} \\
M.J.~Slater, Phys. Lett. {\bf B403} (1997) 57, {\tt hep-th/9701170}

\bibitem{wdvv}
A. Marshakov, A. Mironov and A. Morozov, Phys. Lett. {\bf B389} (1996)
 43, {\tt hep-th/9607109}; 
Mod. Phys. Lett. {\bf A12} (1997) 773, {\tt hep-th/9701014}; 
{\it More Evidence for the WDVV Equations in $N=2$ SUSY Yang-Mills
 Theories}, {\tt hep-th/9701123}; \\
 G. Bonelli and M. Matone, Phys. Rev. Lett. {\bf 77}
 (1996) 4712, {\tt hep-th/9605090}; \\
G. Bertoldi and M. Matone, Phys. Lett. {\bf B425} (1998) 104, {\tt
 hep-th/9712039} 


\bibitem{scale}
M.~Matone,   Phys. Lett. {\bf B357} (1995) 342, {\tt hep-th/9506102} \\
J.~Sonnenschein, S.~Theisen and S.~Yankielowicz,
Phys. Lett. {\bf B367} (1996) 145, {\tt hep-th/9510129} \\
T.~Eguchi and S.-K.~Yang, Mod. Phys. Lett. {\bf A11} (1996) 131,
{\tt hep-th/9510183} 

\bibitem{RG}
A. Gorskii, A. Marshakov, and A, Mironov, Nucl. Phys. {\bf B527}
(1998) 690, {\tt hep-th/9802007}

\bibitem{EdGoMa}
J. D. Edelstein, M. Mari\~no and J. Mas, Nucl. Phys. {\bf B541} (1999)
671, {\tt hep-th/9805172} \\
J. D. Edelstein, M. G\'omez-Reino and J. Mas, 
{\it Instanton corrections in ${\cal N}=2$ supersymmetric theories
with classical gauge groups and fundamental matter hypermultiplets}, 
{\tt hep-th/9904087}

\bibitem{ArPlSeWi}
P. Argyres, R. Plesser, N.Seiberg and E. Witten, Nucl. Phys. {\bf
B461} (1996) 71, {\tt hep-th/9511154}

\bibitem{EgHoItYa}
T. Eguchi, K. Hori, K. Ito and S.-K. Yang, Nucl. Phys. {\bf 471}
(1996) 430, {\tt hep-th/9603002} \\
T. Eguchi and K. Hori, {\it  N=2 Superconformal  Field Theories in
Four-dimensions and  A-D-E Classification}, {\tt hep-th/9607125}

\end{thebibliography}
\end{document}